\documentclass[a4paper,english,numberwithinsect]{eurocg23}

\usepackage{amsmath, amssymb, amsthm, graphicx, verbatim, hyperref, complexity}
\newtheorem{observation}{Observation}

\title{The Complexity of Intersection Graphs of Lines in Space and Circle Orders}
\author{Jean Cardinal}
\affil{Universit\'e libre de Bruxelles (ULB)\\
  \texttt{jean.cardinal@ulb.be}}
\authorrunning{J. Cardinal}
\ArticleNo{37}

\begin{document}
\maketitle
\sloppy

\begin{abstract}
  We consider the complexity of the recognition problem for two families of combinatorial structures.
  A graph $G=(V,E)$ is said to be an intersection graph of lines in space if every $v\in V$ can be mapped to a straight line $\ell (v)$ in $\mathbb{R}^3$ so that $vw$ is
  an edge in $E$ if and only if $\ell(v)$ and $\ell(w)$ intersect.
  A partially ordered set $(X,\prec)$ is said to be a circle order, or a 2-space-time order, if every $x\in X$ can be mapped to a closed circular disk $C(x)$ so that $y\prec x$ if and only if $C(y)$ is contained in $C(x)$.
  We prove that the recognition problems for intersection graphs of lines and circle orders are both $\exists\mathbb{R}$-complete, hence polynomial-time equivalent to deciding whether a system of polynomial equalities and inequalities has a solution over the reals. The second result addresses an open problem posed by Brightwell and Luczak.
\end{abstract}

\section{Introduction}

The complexity class $\exists\mathbb{R}$ is the set of decision problems that are polynomial-time reducible to deciding the validity of a sentence in the existential theory of the reals.
This class was popularized by Schaefer and Stefankovic~\cite{S09,SS17}, and since then many problems have been classified as $\exists\mathbb{R}$-complete~\cite{C15}.
The class $\exists\mathbb{R}$ contains \NP, hence $\exists\mathbb{R}$-complete problems are \NP-hard.
It is also known that $\exists\mathbb{R}\subseteq \PSPACE$~\cite{canny1988}.

Among known $\exists\mathbb{R}$-complete problems, we find many computational geometry problems in which we are asked to decide whether some combinatorial structure has a geometric realization, hence recognition problems for combinatorial families defined by geometric constructions.
This includes, among others, realizable oriented matroids and order types~\cite{M85,M88}, linkages~\cite{S12}, simultaneously embeddable planar graphs~\cite{CK15,S21a}, and several classes of geometric intersection graphs, including segment intersection graphs~\cite{KM94,M14} and disk intersection graphs~\cite{KM12,MM13}. More recently, other fundamental problems in computational geometry have been shown to be $\exists\mathbb{R}$-complete, such as the art gallery problem~\cite{AAM22}, and several families of geometric packing problems~\cite{AMS20}.

In this paper, we prove the $\exists\mathbb{R}$-completeness of the recognition problem for two families of combinatorial objects: (i) intersection graphs of lines in 3-space, and (ii) circle orders.

\subsection{Lines in space}

Arrangements of lines in 3-space are a classical topic in discrete geometry~\cite{E90,CEGSS96,R17}.
We consider the problem of deciding, given a graph $G$, whether we can we map each of its vertices to a line in $\mathbb{R}^3$ so that two vertices are adjacent if and only if the two corresponding lines intersect. Such graphs will be referred to as \emph{intersection graphs of lines in space}.

Pach, Tardos, and T\'oth studied complements of intersection graphs of lines in space, and proved that graphs of maximum degree three and line graphs (adjacency graphs of edges of a graph) are intersection graphs of lines in space~\cite{PTT21}. They also considered the computational complexity of various combinatorial optimization problems on these graphs.
Davies~\cite{D21} showed that intersection graphs of lines in space are not $\chi$-bounded: There exist such graphs that have arbitrary large girth and chromatic number.  
On the other hand, Cardinal, Payne, and Solomon showed that they satisfy a nice Erd\H{o}s-Hajnal property: They either have a clique or an independent set of size $\Omega(n^{1/3})$~\cite{CPS16}.

Regarding the recognition problem, two closely related results to ours have been proved.
First, Matou\u{s}ek and Kratochv\'il showed that intersection graphs of line segments in the plane are $\exists\mathbb{R}$-complete~\cite{KM94}.
This also holds for the more restricted classes of outersegment, grounded segments, and ray intersection graphs~\cite{CFMTV18}.
Evans, Rzazewski, Saeedi, Shin, and Wolff~\cite{ERSS019} proved that intersection graphs of line segments in $\mathbb{R}^3$ are $\exists\mathbb{R}$-complete as well.
We prove that we only need lines for this to hold. 

\begin{theorem}\label{thm:lineIntersection}
  Deciding whether a graph is an intersection graph of lines in space is $\exists\mathbb{R}$-complete. 
\end{theorem}

\subsection{Geometric containement orders}

\emph{Geometric containment orders} are posets of the form $(X,\subset)$, where $X$ is a countable set of (usually convex) subsets of $\mathbb{R}^d$, for some constant $d$, and $\subset$ is the usual containment relation.
Geometric containment orders are well-studied~\cite{FT98}.
The classical \emph{Dushnik-Miller dimension} of a poset can be defined in terms of geometric containment orders, as the smallest $d$ such that the poset is isomorphic to a containment order of translates of the negative orthant in $\mathbb{R}^d$~\cite{T92}.
\emph{Circle orders} are containment orders of closed disks in the plane, and more generally, \emph{$d$-sphere orders} are containment orders of closed balls in $\mathbb{R}^d$~\cite{SSU88,F88,BW89}.
The corresponding notion of dimension is the \emph{Minkowski dimension}~\cite{M93}.

The definition of $d$-sphere orders is motivated by causal set theory in physics and the notion of space-time order~\cite{BG91,RS99,D13,S19}.
In 1992, Scheinerman~\cite{S92} proved that circle orders are also one-to-one with parabola orders and Loewner orders for $2\times 2$ Hermitian matrices, and gave a general equivalence statement for \emph{$Q/L$ orders}, an algebraically defined class of posets. 

In a survey on the combinatorics of the causal set approach to quantum gravity~\cite{BL16}, Brightwell and Luczak pose the problem of recognizing circle orders in those terms: ``It is remarkable that this concrete question is apparently not easy to answer computationally; to the best of our knowledge, the complexity of determining whether a given finite partial order is (for instance) a circle order is unknown.''
We prove that the problem is $\exists\mathbb{R}$-complete.
We recall that a poset is \emph{bipartite} if it can be partitioned into two antichains, or subsets of pairwise incomparable elements.

\begin{theorem}\label{thm:circle}
  Deciding whether a poset is a circle order is $\exists\mathbb{R}$-complete, even when the input is restricted to bipartite posets.
\end{theorem}

\noindent{\bf Outline.}
Theorem~\ref{thm:lineIntersection} is proved in Section~\ref{sec:igl}, and Theorem~\ref{thm:circle} is proved in Section~\ref{sec:co}.
Our proofs build on the classical complexity-theoretic interpretation of Mn\"ev's Universality Theorem~\cite{M85,M88}, and on recent hardness results from Kang and M\"uller~\cite{KM14}, and Felsner and Scheucher~\cite{FS18}.
Note that in both cases, containment in $\exists\mathbb{R}$ is easy to prove, and we concentrate on the hardness part.

\noindent{\bf Acknowledgments.}
The proof of Theorem~\ref{thm:lineIntersection} is based on an idea due to Udo Hoffmann.
The author wishes to thank Till Miltzow and the EuroCG referees for their comments on a preliminary version.

\newpage
\section{$\exists\mathbb{R}$-Completeness of intersection graphs of lines}
\label{sec:igl}

Our proof is by reduction from the affine realizability problem for matroids of rank 3.
In short, we are given a ground set $E$ together with a collection $\mathcal{I}$ of subsets of $E$
defining the maximal independent sets of the matroid. In rank 3, the maximal independent sets have size three, hence $\mathcal{I}$ is a collection of triples.
We wish to decide whether there exists a map $f:E\to\mathbb{R}^2$ such that for every triple $a,b,c\in E$, the points $f(a),f(b),f(c)$ form a nondegenerate triangle if and only if $\{a,b,c\}\in\mathcal{I}$.
It is somehow a folklore result that this problem is $\exists\mathbb{R}$-complete, and a detailed proof was given recently by Kim, de Mesmay, and Miltzow~\cite{KMM23}.

Given a graph $G=(V,E)$, a \emph{line realization} of $G$ is a map $\ell$ from $V$ to the set of lines in $\mathbb{R}^3$ such that $vw\in E\Leftrightarrow \ell(v)\cap\ell(w)\not=\emptyset$.
For a subset $S\subseteq V$, we use the notation $\ell(S)=\{\ell(v):v\in S\}$.
The following observation will be useful.

\begin{observation}\label{obs:cliquePlane}
  In a line realization of a complete graph on the vertex set $V$, the lines of $\ell(V)$ intersect in one point or are coplanar.
\end{observation}

We also use the following statement, which allows us to force a collection of lines in a line realization to lie in a plane.

\begin{lemma}\label{lem:forceInPlane}
  Let $G_n$ be the complete graph on $2n$ vertices minus a perfect matching.
  For $n>3$ all lines of a line realization $\ell$ of $G_n$ are coplanar.
\end{lemma}
\begin{proof}
  Let $(A,B)$ be the bipartition of the vertices of $G_n$ into two cliques of size $n$, and consider a line realization $\ell$ of $G_n$.
  We show that the lines of $\ell(A)$ are coplanar.
  For contradiction, suppose otherwise.
  Then from Observation~\ref{obs:cliquePlane}, all lines of $A$ must intersect in one point $p$.
  Furthermore, there are three vertices $a_1,a_2,a_3\in A$ such that the lines $\ell(a_1), \ell(a_2), \ell(a_3)\in \ell(A)$ are not coplanar.
  There is also a vertex $b\in B$ adjacent to $a_1,a_2$ and $a_3$ in $G_n$, but not adjacent to a fourth vertex $a_4\in A$. 
  Hence the line $\ell(b)$ must intersect $\ell(a_1), \ell(a_2)$ and $\ell(a_3)$ but avoid $\ell(a_4)$.
  The only possibility to intersect all three lines is in $p$, but then $\ell(b)$ also intersects $\ell(a_4)$, a contradiction.
  By symmetry, all lines in $\ell(B)$ must be coplanar as well, and lying in the same plane as those in $\ell(A)$.
\end{proof}

\begin{figure}[h]
  \begin{center}
    \includegraphics[page=5, width=.9\textwidth]{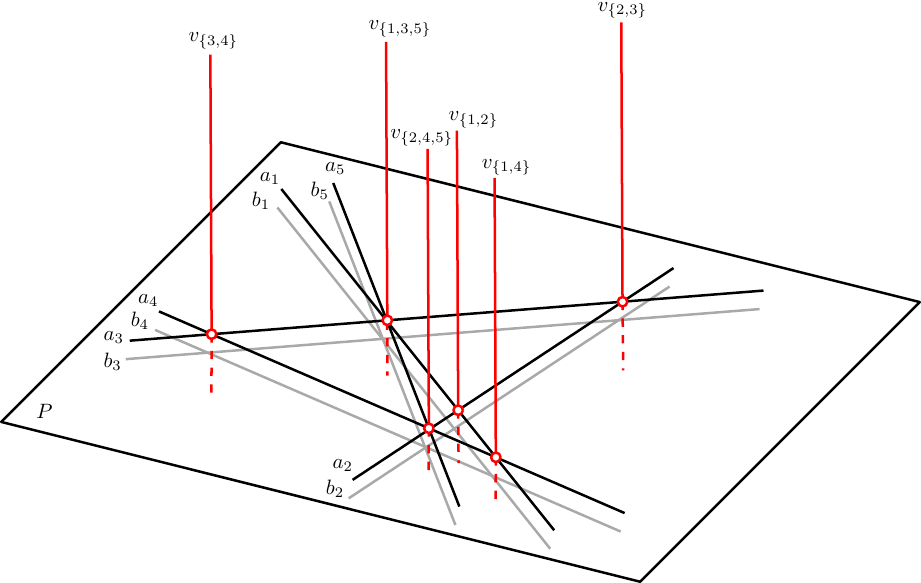}
  \end{center}
  \caption{\label{fig:matroid}A line realization of the graph $H$ constructed from a realizable rank-3 matroid on 5 elements
    with rank-2 flats $\mathcal{S}=\{\{1,2\}, \{2,3\}, \{3,4\}, \{1,4\}, \{1,3,5\}, \{2,4,5\}\}$.}
  \end{figure}

\begin{proof}[Proof of Theorem~\ref{thm:lineIntersection}]
  We reduce from realizability of a rank-3 matroid $M=(E,\mathcal{I})$, as defined above.
  It is convenient to let $E=\{1,2,\ldots ,n\}$, and consider the set system $\mathcal{S}$ on $E$ defined as the set of rank-2 flats of $M$.
  In a realization of $M$, these flats correspond to all inclusionwise maximal subsets of collinear points.
  Note that $|\mathcal{S}|\leq {n\choose 2}$.
  
  We start by considering a line realization of the graph $G_n$ defined in Lemma~\ref{lem:forceInPlane}.
  We consider the bipartition of the vertices of $G_n$ into the two cliques $A=\{a_1,\dots,a_n\}$ and $B=\{b_1,\dots,b_n\}$, where $a_i$ is adjacent to $b_j$ if $i\not=j$.
  Then for each set $C\in \mathcal{S}$, we add a new vertex $v_C$ and an edge $v_Ca_i$ for each $i\in C$.
  We let $H$ be the resulting graph, defined on the vertex set $A\cup B\cup \{v_C:C\in\mathcal{S}\}$.
  We claim that $H$ is a intersection graph of lines if and only if the input matroid $M$ is realizable.

  To construct a realization of $M$ from a line realization $\ell$ of $H$ we use planar point-line duality on the lines in $\ell(A)$.
  From Lemma~\ref{lem:forceInPlane}, the lines of $\ell(A\cup B)$ all lie in a plane $P$.
  For every set $C\in \mathcal{S}$, the vertices $\{v_C\}\cup\{a_i: i\in C\}$ form a maximal clique in $H$.
  From Observation~\ref{obs:cliquePlane}, this clique must be realized by an intersection point common to all lines $\{\ell(a_i): i\in C\}$, since otherwise $\ell(v_C)$ lies in $P$ and intersects some lines in $\ell(B)$.
  Point-line duality in $P$ maps lines with a common intersection point to collinear points.
  We denote by $p(a_i)$ the point that is dual to the line $\ell(a_i)$ in $P$.
  Therefore, the dual points $\{p(a_i): i\in C\}$ are collinear in $P$.
  From the maximality of the clique $\{v_C\}\cup\{a_i: i\in C\}$, the points $\{p(a_i): i\in C\}$ form a maximal collinear set, and $p(A)$ is a realization of $M$ in $P$.

  On the other hand, a realization of $M$ in the plane can be mapped to a dual arrangement of lines.
  Let $A=\{a_i:i\in E\}$, and $\ell(a_i)$ be the line representing element $i$, for every $i\in E$.
  Now each line $\ell(b_j)\in \ell(B)$ can be represented by a line parallel to $\ell(a_j)$ and intersecting all the other lines $\ell(a_i)$ and $\ell(b_i)$ for $i\not= j$.
  We embed this realization into a plane $P$ in $\mathbb{R}^3$ and add a line $\ell(v_C)$ for each $C\in \mathcal{S}$ that pierces $P$ in the common intersection point of the lines $\{a_i: i\in C\}$ (see Figure~\ref{fig:matroid}).
  This yields a line realization of $H$.
\end{proof}

\section{$\exists\mathbb{R}$-Completeness of circle orders}
\label{sec:co}

\begin{figure}[h!]
  \begin{center}
    \includegraphics[page=3, width=.4\textwidth]{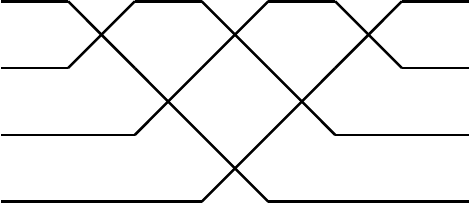}
    \hspace{1cm}
    \includegraphics[page=4, width=.4\textwidth]{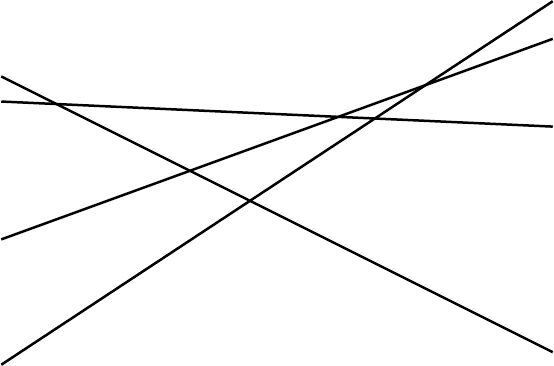}
  \end{center}
  \caption{\label{fig:lines}A pseudoline arrangement and an isomorphic line arrangement.}
  \end{figure}

A \emph{pseudoline arrangement} is a collection of $x$-monotone curves that pairwise intersect exactly once.
A pseudoline arrangement is \emph{simple} if no triple of lines intersect in the same point, and \emph{stretchable} if it is isomorphic to an arrangement of straight lines, see Figure~\ref{fig:lines}.
A well-known consequence of Mn\"ev's Theorem -- see Shor~\cite{Sh91}, and Richter-Gebert~\cite{RG95} -- is that deciding whether a given pseudoline arrangement is stretchable is $\exists\mathbb{R}$-complete.

\begin{figure}[h!]
  \begin{center}
  \includegraphics[page=1, width=.20\textwidth]{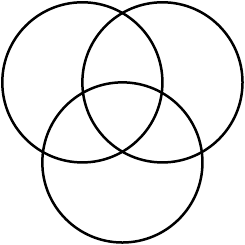}
    \hspace{1cm}
  \includegraphics[page=2, width=.4\textwidth]{circleorders_figs.pdf}
  \end{center}
  \caption{\label{fig:gpca}A Krupp, and a great-pseudocircle arrangement realizing two copies of the pseudoline arrangement of Figure~\ref{fig:lines}.}
  \end{figure}

In order to deal with circle orders, we make use of recent analogous results on arrangement of circles due to Kang and M\"uller~\cite{KM14}, and Felsner and Scheucher~\cite{FS18}.
An arrangement of Jordan curves is said to be a \emph{pseudocircle arrangement} if every pair of curves is either disjoint or intersects in exactly two points.
A triple of pseudocircles is said to form a \emph{Krupp} if they pairwise intersect in distinct points and  their interiors have a common point.
A pseudocircle arrangement is said to be a \emph{great-pseudocircle arrangement} if every triple of pseudocircles forms a Krupp. See Figure~\ref{fig:gpca} for illustrations.
Similarly to pseudoline arrangements, we are only concerned with the topological structure of the arrangement.
Great-pseudocircle arrangements constitute a combinatorial abstraction of arrangements of great circles on a sphere, hence of line arrangements in the projective plane.
In fact, great-pseudocircle arrangements and simple pseudoline arrangements are essentially the same combinatorial family, see Felsner and Scheucher~\cite{FS18}.

\begin{lemma}
  Great-pseudocircle arrangements are one-to-one with simple pseudoline arrangements.
\end{lemma}

Every great-pseudocircle arrangement realizes two copies of a pseudoline arrangement, see Figure~\ref{fig:gpca}.
When stretchable, they correspond to realizations of pseudolines as great circles on the 2-sphere, or equivalently as lines in the projective plane.

We need a technical lemma on the structure of circle arrangements.
A collection of $n$ circles in the plane partitions the plane into connected subsets that we refer to as \emph{cells}.
To each cell, one can assign a binary string of length $n$ indicating, for each circle, whether the cell lies in the interior or the exterior of the circle.
We refer to this string as the \emph{label} of the cell.

\begin{lemma}
  \label{lem:count}
  The maximum number of dictinct cell labels of an arrangement of $n$ circles is equal to the maximum number of cells of a circle arrangement, and is $n(n-1)+2$.
  This bound is attained by great-pseudocircle arrangements.
\end{lemma}
\begin{proof}
  One can proceed by induction on $n$. We let $f(n)$ be the maximum number of labels, with $f(1)=2$.
  When adding the $n$th circle to the arrangement, the number of additional labels is at most the number of newly created cells.
  This number is in turn bounded by the number of new intersection points:
  One can charge every new cell to the intersection point on its left in a clockwise sweep around the new circle.
  Therefore $f(n)\leq f(n-1)+ 2(n-1)$, solving to the announced bound.
  From this reasoning, tight examples must be such that no two cells have the same label, and one can check that this bound is attained by great-pseudocircle arrangements.
\end{proof}

The \emph{circularizability} problem is the problem of deciding whether a pseudocircle arrangement is isomorphic to a circle arrangement.
It is the analogue of the stretchability problem for pseudoline arrangements, and is in fact polynomial-time equivalent to it.

\begin{theorem}[Kang and M\"uller~\cite{KM14}, Felsner and Scheucher~\cite{FS18}]
  Deciding whether a pseudocircle arrangement is circularizable is $\exists\mathbb{R}$-complete, even when the input is restricted to great-pseudocircle arrangements.
\end{theorem}

With these tools at hand, we now prove our second main result.

\begin{proof}[Proof of Theorem~\ref{thm:circle}]
  We reduce from the circularizability problem.
  Given a great-pseudocircle arrangement $\mathcal{C}$, we construct a bipartite poset $P=(X,\prec)$ that is a circle order if and only if $\mathcal{C}$ is circularizable.
  We define the ground set $X$ of $P$ as $X:=C\cup F$, where $C$ is the set of great-pseudocircles of $\mathcal{C}$, and $F$ is the set of cell labels of $\mathcal{C}$.
  Both $C$ and $F$ induce empty posets in $P$, and for a pair $(c,f)\in C\times F$, we let $c\prec f$ if and only if the cell corresponding to $f$ lies in the disk bounded by $c$.
  We show that $P$ is a circle order if and only if $\mathcal{C}$ is circularizable.
  
  Indeed, if $\mathcal{C}$ is circularizable, then we can consider a circle realization of $\mathcal{C}$ and add a small circle within each cell of the arrangement to realize $P$.
  For the other direction, suppose $P$ has a realization as a circle order and consider the arrangement $\mathcal{D}$ induced by $C$ in this realization.
  Since every $f\in F$ has an associated circle as well, every cell label of $\mathcal{C}$ is realized in $\mathcal{D}$. From Lemma~\ref{lem:count}, the number of cells is as large as can be for an arrangement of circles, hence no other cell can appear in $\mathcal{D}$. Therefore $\mathcal{D}$ must be a circularization of $\mathcal{C}$.
  \end{proof}

Note that a similar proof was given by Tanenbaum, Goodrich, and Scheinerman~\cite{TGS94} to prove hardness of point-halfspace incidence orders.
Also note that from a theorem of Scheinerman~\cite{S91}, the vertex-edge incidence poset of a graph is a circle order if and only if the graph is planar.
Hence for vertex-edge incidence posets, the problem reduces to planarity testing and can therefore be solved in linear time. 


\bibliographystyle{plain}
\bibliography{circleorders}

\end{document}